
\input phyzzx.tex
\PHYSREV
\pubnum{ROM2F-93-03}

\def\inte{ {1\over\pi}\int d^2\sigma\ }
\def\de{\partial}
\def\kin#1{\de_+ #1 \de_- #1}
\def\primo{^\prime}
\def\intx{{1\over 2\pi}\int d^2 x\ }
\def\kinx#1{\de_\mu #1\de^\mu #1}
\def\curv#1{i q_#1 R #1}
\def\unm{{1\over 2}}
\def\dmu#1{\mu{d #1\over d\mu}}
\def\bettyr#1{\beta_#1 := \dmu{#1^2_R}}
\def\boopr{\unm\gamma_{+R}\gamma_{-R}}

\def\lm{{\lambda^{(-)}}}

\def\radgR{\sqrt{-\hat g}\hat R}
\def\xop{x_0^+}
\def\mbh{m_{\rm bh}}
\def\uq{U_q(\widetilde{sl(2)})}
\def\al{\vec\alpha}
\def\tphi{\tau_\varphi}
\def\teta{\tau_\eta}
\def\Qb{\bar Q}
\def\stilde{\tilde S}

\def\sbh{\stilde_{\rm bh}}
\titlepage
\title{A CONFORMAL AFFINE TODA MODEL OF 2D-BLACK HOLES\break
THE END-POINT STATE AND THE S-MATRIX}
\author{F.~Belgiorno, A.~S.~Cattaneo\foot{Also Sezione
I.N.F.N. dell'Universit\`a di Milano, 20133 Milano, Italy}
}
\address{Dipartimento di Fisica, Universit\`a di Milano,
20133 Milano, Italy}
\author{F.~Fucito}
\address{Dipartimento di Fisica, Universit\`a di Roma II ``Tor Vergata"
and INFN sez. di Roma II,00173 Roma, Italy}
\andauthor{ M.~Martellini\foot{
Permanent address: Dipartimento di Fisica, Universit\`a di Milano, 20133
Milano, Italy and
also Sezione I.N.F.N.
dell'Universit\`a di Pavia, 27100 Pavia, Italy}}
\address{Dipartimento di Fisica, Universit\`a di Roma I ``La Sapienza",
00185 Roma, Italy}
\abstract
{In this paper we investigate in more detail our previous formulation of
the dilaton-gravity theory by Bilal--Callan--de~Alwis as
a $SL_2$-conformal affine Toda (CAT) theory. Our main results are:
\item {i)} a field redefinition of the CAT-basis in terms of which it is
possible to get the black hole solutions already known in the literature;
\item {ii)} an investigation the scattering matrix problem for the
quantum black hole states.
\par
It turns out that there is a range of values of the $N$ free-falling
shock matter fields forming the black hole solution, in which the
end-point state of the black hole evaporation is a zero temperature
regular remnant geometry. It seems that the quantum evolution to this
final state is non-unitary, in agreement with Hawking's scenario
for the black hole evaporation.}
\endpage
\pagenumber=1
\chapter{Introduction}
\REF\CGHS{C.~G.~Callan, S.~B.~Giddings, J.~A.~Harvey, and A.~Strominger
\journal Phys. Rev. &D45(92)R1005.}
\REF\BDD{T.~Banks, A.~Dabholkar, M.~R.~Douglas, and M.~O'Loughlin
\journal Phys. Rev. &D45(92)3607.}
\REF\RST{J.~G.~Russo, L.~Susskind, and L.~Thorlacius,
{\it Black Hole Evaporation in $1+1$ Dimensions},
Stanford preprint SU-ITP-92-4, January 1992;\nextline
L.~Susskind and L.~Thorlacius,
{\it Hawking Radiation and Back-Reaction},
Stanford preprint SU-ITP-92-12, hepth@xxx/9203054, March 1992.}
\REF\BGHS{B.~Birnir, S.~B.~Giddings, J.~A.~Harvey and A.~Strominger,
{\it Quantum Black Holes}, Santa Barbara preprint UCSB-TH-92-08,
hepth@xxx/9203042, March 1992.}
\REF\Str{A.~Strominger, {\it Faddeev--Popov Ghosts and $1+1$ Dimensional
Black Hole Evaporation},
UC Santa Barbara preprint UCSB-TH-92-18, hept@xxx/9205028, May 1992.}
\REF\Haw{S.~W.~Hawking, {\it Evaporation of Two Dimensional Black Holes},
CALTECH preprint CALT-68-1774, March 1992.}
In a series of recent papers\refmark\CGHS$^{^-}$\refmark\Haw
a dilaton-gravity $2D$-theory which has
black hole solutions, known as the Callan--Giddings--Harvey--Strominger
(CGHS) model, was formulated in order to clarify the problem of the
black hole evaporation due to Hawking radiation.
Later Bilal and
Callan\Ref\BC{A.~Bilal and C.~Callan, {\it Liouville Models of Black Hole
Evaporation}, Princeton preprint PUPT-1320, hept@xxx/9205089, May 1992.}
and de~Alwis\Ref\deA{S.~P.~de~Alwis, {\it Black Hole Physics from
Liouville Theory}, Colorado preprint COLO-HEP-284, hepth@xxx/9206020,
June 1992.}
have reformulated the CGHS-model as a
Liouville-like field theory, so that one may obtain some exact results using
just Liouville-like techniques.
\par
However the resulting theory, which we shall call the Liouville-like
black hole (LBH) theory, is still ill-defined when the string coupling
constant $e^{2\phi}$ ($\phi$ is the dilaton) is of the order of a certain
critical value ${1\over\kappa}$, with $\kappa$ the coefficient of the
Polyakov kinetic term in the LBH-model. In this limit a singularity must
occur.
Furthermore the CGHS and LBH models are consistent if one could neglect
graviton-dilaton quantum effects. This amounts to require a ``large"
number $N$ of conformal invariant matter fields interacting with the black
hole geometry. In the LBH-model this condition implies large positive
$\kappa$, since $\kappa={N-24\over 12}$. At finite $N$ orders,
quantum loop effects of the graviton and dilaton are important and their
resulting effective $2D$-geometry may be totally different from the one
derived by the LBH-model, say for $N\buildrel <\over\sim 24$
($\kappa\buildrel <\over\sim 0$).
\par
In this paper we extend the results of our previous
analysis\rlap,\Ref\BCFM{F. Belgiorno
A.S. Cattaneo, F. Fucito and M. Martellini,
{\it A Conformal Affine Toda Model of 2D Black Holes: A Quantum Study of the
Evaporation End-Point}, ROM2F 92/52, October 1992,\nextline
F. Belgiorno, A.S. Cattaneo, F. Fucito and M. Martellini, {\it Quantum
Models of Black-Hole Evaporation} in {\sl Proceedings of the Roma
Conference on ``String Theory, Quantum Gravity and the Unification of
Fundamental Interactions"}, preprint ROM2F 92/60, November 1992.}
in which we showed that
when $e^{2\phi}\sim{1\over\kappa}$, the LBH-model can be reformulated
as a conformally invariant integrable
theory based on the $SL_2$-Kac--Moody algebra, which is known as the
conformal affine Toda (CAT) theory\rlap.\Ref\BB{O.~Babelon
and L.~Bonora \journal Phys. Lett. &B244(90)220;\nextline
O. Babelon and L. Bonora, {\sl Phys. Lett.}\/ {\bf B267(1991),71.}
}
In the following we shall call our $2D$ black-hole model
the conformal affine Toda black hole (CATBH).
The CATBH-model allows a standard perturbative quantization
and in our picture such a quantization is just a device to unveal the
quantum effect of the dilaton $\phi$ and graviton $\rho$ because the
CAT-fields are suitable functions of ($\phi,\rho$).
\par
This full quantum
reformulation of the $2D$-black-hole has several
advantages:
\item{i)} it makes possible an investigation of the physics around the
CGHS singularity, using the conformal field theory associated to
the CAT-model (Sec. 2). Furthermore, by using a suitable ``rotation" of
the CAT-fields it is possible to map our equations of motion to those
given in Ref. \BC, which exhibit the well-known $2D$-black-hole solution.
\item{ii)} we can study the final state of the black hole evaporation by
applying the Renormalization Group (RG) analysis to the CATBH-model.
The model has non-trivial fixed points and an energy scale where it reduces to
the LBH-model, where we can get effective
dynamical relations (with respect to the RG-scale)
for the Hawking temperature and for the v.~e.~v. of the
$2D$-density scalar curvature in the tree approximation.
The RG-analysis is given in Sec. 3.
\item{iii)} Starting from the CAT-model, which is quantum integrable
it is possible to guess (Sec. 5) a non-trivial $S$-matrix
for the quantum black-hole states.
\par
The back-reaction should modify the Hawking radiation emission
and cause it to stop when the black hole has radiated away its initial ADM
mass. In our context the Hawking temperature is proportional to the
square root of a certain
coupling constant, $\gamma_-$, of an exponential interaction term of the
CAT-model, which now may be regarded (by the RG analysis of Sec. 3)
as a running
coupling constant in terms of the energy-mass scale $\mu$, which roughly
speaking measures the ADM mass. Similarly, one finds that the averaged
curvature in the tree approximation and in the so-called conformal vacuum
may be also related to $\gamma_-$.
As a consequence
one knows how the strength of the effective curvature varies as a function
of the Hawking temperature for fixed $\mu$. In particular if we formulate
an ansatz for the black hole solution interacting with $N$ in-falling
free shock waves,
one may study the thermodynamics of the end point of the black hole evaporation
by extrapolating the RG-effective couplings to the (classical) energy scale
$\hat t$ where the initial ADM-mass goes to zero. This picture gives an
approximate self consistent scheme to take into account the back reaction
problem in the mechanism of the Hawking radiation: the back reaction lies
in a dependence of the Hawking temperature over an energy scale $t=\log\mu$
and the
end point of the black hole evaporation is characterized in the
CATBH-model by the limit
$t\to\hat t$ where $M(t)\to M(\hat t)\sim 0$. Here $M(t)$ is the classical
ADM-mass parametrized by $t$ and $M(\hat t)\sim 0$
describes the end point state in which all the initial ADM-mass has been
radiated away. We shall show in Sec. 4 that the end-point state of the
black-hole evaporation of a $2D$-CGHS black-hole interacting with
$N$ shock waves, when $N\in(16,23)$, is characterized by a
zero-temperature regular remnant geometry.
\par
In the last section, Sec. 5, we shall use the quantum
integrability of the CATBH model to extract the universal
$R$-matrix, which turns out to be associated
to the quantized affine centrally extended $\widetilde{sl_2}$-algebra.
Regarding the $R$-matrix (up to a function)
as a two body $S$-matrix operator, one gets the evolution
operator between the states which describe the full quantum $2D$-black
hole in the conformal affine Toda basis.
Pulling back such a CAT $S$-matrix to the ``physical"
black hole basis described by the
states $\ket{\chi,u=v}$, associated to the fields $\chi,u,v$ of Sec. 2,
one obtains the
$S$-matrix operator $\sbh$ corresponding to the quantum evolution of the
black-hole geometry. It results that $\sbh$ is not unitary, thus
confirming Hawking's
scenario\Ref\Hawk{S.W. Hawking \journal Phys. Rev. &D14(76)2460.}
for the
black hole evaporation and the loss of
quantum coherence.
\chapter{Conformal Affine Toda Black Hole Model}
Bilal  and Callan\refmark\BC\
have recently shown how to cast the CGHS black hole model
in a non standard Liouville-like form. Their key idea is to mimic
the Distler--Kawai\Ref\DK{J.~Distler and H.~Kawai \journal Nucl. Phys.
&B321(89)509.}
approach to $2D$-quantum gravity and include in the action
a dilaton dependent renormalization of the
cosmological constant as well as a Polyakov effective term induced by the
``total" conformal anomaly. The effect of the Polyakov term
in the ``flat" conformal gauge $\widehat{g_{\mu\nu}}
=-\unm e^{2\rho}\eta_{\mu\nu}$ will be to replace the coefficient ${N\over 12}$
of CGHS by a coefficient $\kappa$, where $\kappa$ will be fixed by
making the matter central charge (due to the $N$ free
scalar fields) cancel against the $c=-26$ diffeomorphism-ghosts contribution.
More specifically they were able to simplify the graviton-dilaton
action through a field redefinition of the form
$$
\matrix{
\omega = {e^{-\phi}\over \sqrt{\kappa}},
&\hfill
& \chi = {1\over 2} (\rho + \omega^2)},
\eqn\redoc
$$
where $\phi$ and $\rho$ are respectively the dilaton and the Liouville mode
and, in the LBH-model, $\kappa = {N-24\over 12}$  as it follows by the
requirement of conformal invariance.
According to our way of thinking $\kappa$ must now be a free parameter.
However for  positive $\kappa$
a singularity appears in the regime $\omega^2\to 1$,
which is also a strong coupling limit for small $\kappa$.
\par
Our purpose, as stated in the introduction, is to define an improvement of
the Bilal--Callan theory that allows for a well defined exact conformal
field theory to exists also in the  positive $\kappa$ ``phase"and
$\omega^2\sim 1$.
The surprise is that such a theory must be a sort of ``massive
deformation" (which must be conformal at the same time)
of the true Liouville theory,
as we shall see in the following.
\par
The kinetic part of the Bilal--Callan action is given by
$$
S_{\rm kin} = \inte\left[ {-4\kappa\kin{\chi} + 4\kappa (\omega^2 -1)
\kin{\omega}+{1\over 2} \sum_{i=1}^N \kin{f_i} }\right] ,
\eqn\bc
$$
and the energy-momentum tensor has the Feigin--Fuchs form
$$
T_\pm=-4\kappa\de_\pm\chi\de_\pm\chi+2\kappa\de^2_\pm\chi
+4\kappa(\omega^2-1)\de_\pm\omega
\de_\pm\omega+{1\over 2}\sum_{i=1}^N\de_\pm f_i\de_\pm f_i.
\eqn\bcst
$$
\par
When $\omega^2 >1$ the action and the stress tensor can be simplified
by setting
$$
\Omega = {1\over 2}\omega\sqrt{\omega^2-1}-
{1\over 2}\log{(\omega +\sqrt{\omega^2-1})} ,
\eqn\defom
$$
leading to the canonical kinetic term $4\kappa\kin{\Omega}$.
\par
When $\omega^2 <1$ we must use the alternative definition
$$
\Omega\primo = {1\over 2}\omega\sqrt{1-\omega^2}-{1\over 2} \arccos{\omega} ,
\eqn\defompr
$$
obtaining the ``ghost-like" kinetic term $-4\kappa\kin{\Omega\primo}$.
\par
We thus have two different theories describing different regions of the
space-time (remember that $\omega$ is a field), in the first one we have
a real positive valued field $\Omega$,
in the latter a ghost field $\Omega\primo$
with range in $[-{\pi\over 4},0]$. The first theory is the one we are
interested
in when considering the classical limit $\omega\to +\infty$
and has been studied in Ref. \BC.
If we choose not to constrain the range of values of $\omega$, the
quantization of the LBH theory (the one in which $\omega\in [1,
\infty]$) should provide an IR effective theory of the complete one
in which $\omega$ has its natural range from $0$ to $\infty$.
In other words, in a full quantized theory one cannot forget that somewhere the
field $\omega (\sigma )$ can take values less than 1,
in which case the ghost-like action form must be used.
\par
Particularly interesting is the region where $\omega^2(\sigma)-1$
is changing sign. Let us suppose that $\omega^2(\sigma_1)
> 1$ and $\omega^2(\sigma_2) < 1$, where $\sigma_1$ and
$\sigma_2$ are two very close points. Then our idea is to define an
improved Bilal--Callan kinetic action term in which both fields
$\Omega(\sigma_1)$ and $\Omega\primo(\sigma_2)$ appear. Namely we assume
the following contribution to such an improved action:
$$
4\kappa(\kin{\Omega(\sigma_1)}-\kin{\Omega\primo(\sigma_2)})
= 2 (\de_+ u(\sigma)\de_- v(\sigma) +
\de_-u(\sigma)\de_+v(\sigma) ),
$$
where the new fields
$$
\eqalign {
u(\sigma) &= \sqrt\kappa(\Omega(\sigma_1) + \Omega\primo(\sigma_2)),\cr
v(\sigma) &= \sqrt\kappa(\Omega(\sigma_1) - \Omega\primo(\sigma_2)),\cr }
\eqn\defex
$$
are defined in the point $\sigma = {\sigma_1 + \sigma_2\over 2}$.
\par
Let us now consider a field $\omega(\sigma)$ taking
values everywhere near 1
and rapidly fluctuating. We can
renormalize (\'a la Wilson) the above theory in which the undefined
Bilal--Callan kinetic term
$(\omega^2-1)\kin{\omega}$ has been replaced by the undefined kinetic term
$\de_+ u\de_- v +\de_- u\de_+ v$. From a naive dimensional argument
when $\omega^2\sim 1$ the Laplacian term $\kin\omega$ must be very large
in order to have a non trivial propagator for the $\omega$ field.
This means that our theory is a sort of ``UV effective theory" for the LBH
model.
It is to be noted that the new fields $ v$ and $ u$ are limited from
below, but since the region of interest is around 0 this constraint is
significative only for $ u$, which has to be positive.
\par
The kinetic part of the ``averaged" LBH model has now the form:
$$
S=\intx\left({
-\unm\kinx{\bar\chi}+\de_\mu u\de^\mu v}\right),
\eqn\save
$$
where we have come back to the coordinates $x_0=(\sigma_++\sigma_-)/2$
and $x_1=(\sigma_+-\sigma_-)/2$ and $\bar\chi=\sqrt{2\kappa}\chi$. We
then take a potential term of the form
$$
V_-=\gamma_-e^{\sqrt{8\over\kappa}\left({
\bar\chi-{u+v\over \sqrt2}}\right)},
\eqn\Vminus
$$
so that the equations of motion take the form:
$$
\eqalign{
\de_\mu\de^\mu\bar\chi &=-\gamma_-\sqrt{8\over\kappa}
e^{\sqrt{8\over\kappa}\left({
\bar\chi-{u+v\over \sqrt2}}\right)},\cr
\de_\mu\de^\mu u=\de_\mu\de^\mu v &=-{\gamma_-\over\sqrt2}\sqrt{8\over\kappa}
e^{\sqrt{8\over\kappa}\left({
\bar\chi-{u+v\over \sqrt2}}\right)}.\cr}
\eqn\BCeq
$$
Notice that there is a class of solutions in which $u=v$, which is
equivalent to $\Omega'=0$ and to $\omega^2>1$. This is the class we are
interested in when considering classical solutions with boundary
conditions corresponding to the linear dilaton vacuum, where
$\omega^2\to+\infty$. Putting $u=v=\sqrt\kappa\Omega$ in \BCeq, as
required by \defex, we obtain
the equations of motion of the LBH model. The coupling constant $\gamma_-$
can now be recognized, apart from a multiplicative factor, as the square
of the cosmological constant.
\par
We can further handle the action \save\ considering a ``rotation" $G$ of the
fields
$$
\left(\matrix{
\bar\chi\cr u\cr v\cr}\right) =
G\cdot\left(\matrix{\varphi\cr\xi\cr\eta\cr}\right),
\eqn\rot
$$
which keeps the kinetic term unchanged, \ie\ in terms of
$\varphi,\xi,\eta$:
$$
S=\intx\left({
-\unm\kinx\varphi+\de_\mu\xi\de^\mu\eta}\right).
\eqn\skin
$$
$G$ is explicitly given by
$$
G=1+A\sinh\sqrt{2ab+c^2}+A^2(\cosh\sqrt{2ab+c^2}-1),
\eqn\defG
$$
where
$$
A={1\over\sqrt{2ab+c^2}}\left(\matrix{
0 & b & a\cr
a & c & 0\cr
b & 0 &-c\cr}\right).
\eqn\defA
$$
$a,b,c$ are three free parameters which will
be fixed by the following three conditions:
\item{i)} we require the potential $V_-$ to take the form
$$
V_-=\gamma_-e^{\lambda\varphi-\delta\eta},
\eqn\vminus
$$
where $\lambda$ and $\delta$ are to be fixed by the requirement that
$V_-$ should have conformal weight (1,1), with respect to the stress
tensor of the theory\rlap.\foot{The stress tensor of the
theory can be calculated by the same procedure
of averaging, \defex. It contains an improved term in $\bar\chi$,
see Ref. \BC, but no terms in $u$ and $v$. After the rotation we have
nevertheless improved terms for all the fields with the background charges
depending on $\kappa$ and $G$.}
\item{ii)}  the theory is invariant under the scale transformations $\xi\to
C\xi,\eta\to\eta/C$, with $C$ arbitrary.
To fix $C$ we further require $\lambda=\delta$.
\item{iii)} to detemine the last degree of freedom in
$G$\rlap.\foot{Notice that since the conditions depend on $\kappa$,
$G$ will depend on it as well.}
we impose the vertex operator
$$
V_+=\gamma_+e^{-\bar\lambda\varphi}
\eqn\Vplus
$$
to be of conformal weight (1,1) and set
$\bar\lambda=\lambda$.
These conditions can be written as equations for the parameters $a,b,c$.
In first place let us notice that the condition $i)$ is equivalent to ask that
$$
\de_\mu\de^\mu u = \de_\mu\de^\mu v.
\eqn\deuv
$$
Furthermore the equations of motion coming from \skin\ with the addition
of the potential term ~\vminus\ (as we shall see in the next section
the vertex \Vplus\ decouples by quantum effect) imply that
$$
{\de_\mu\de^\mu\varphi\over\lambda}={\de_\mu\de^\mu\xi\over\delta}.
\eqn\BBeq
$$
\deuv\ and \BBeq\ give
$$
(G_{21}-G_{31})\lambda=(G_{32}-G_{22})\delta,
\eqn\eqld
$$
where
$$
\eqalign{
\lambda &= \sqrt{8\over\kappa}\left({
G_{11}-{G_{21}+G_{31}\over\sqrt2}}\right),\cr
\delta &= -\sqrt{8\over\kappa}\left({
G_{13}-{G_{23}+G_{33}\over\sqrt2}}\right).\cr}
\eqn\defld
$$
Using condition $ii)$, we get
$$
G_{21}-G_{31}=G_{32}-G_{22},
\eqn\equno
$$
and
$$
G_{11}-{G_{21}+G_{31}\over\sqrt2} =
{G_{23}+G_{33}\over\sqrt2}-G_{13}.
\eqn\eqdue
$$
Condition $iii)$ will be exploited in the next section, eq. (3.31).
\par
We are now in a position to further generalize the LBH model adding to it the
vertex $V_+$, which, by construction, does not alter the conformal
invariance of the theory.
Thus we obtain the conformal affine Toda theory based on
$sl(2)$ proposed by Babelon and Bonora in Ref.~\BB:
$$
S=\intx\left({
-\unm\kinx\varphi+\de_\mu\eta\de^\mu\xi
+\gamma_+e^{-\lambda\varphi}
+\gamma_-e^{\lambda\varphi-\delta\eta}}\right).
\eqn\bblike
$$
Even if this is not the most general action constructed with conformal
perturbations of weight (1,1) (as other exponential combinations of the
fields $\varphi,\xi$ and $\eta$ are possible), we think that it is
actually sufficient.
Indeed very recently Giddings and Strominger\Ref\GS{S.~B.~Giddings and
A.~Strominger, {\it Quantum Theory of Dilaton Gravity}, UC Santa Barbara
preprint, UCSB-TH-92-28, hepth@xxx/9207034, July 1992.}
have argued that there is an infinite number of quantum theories of
dilaton-gravity and the basic problem is to find physical criteria to narrow
the class of solutions. Our approach here is to consider a theory that
\item{i)}
 is at the same time classically integrable, \ie\ admitting a Lax
pair and conformally invariant\rlap,\foot{Notice that \bblike\ is the
unique action coming from the homogeneous gradation of the $\widetilde{sl_2}$
Kac--Moody algebra in terms of three scalar fields.}
\item{ii)}
reduces to the solutions of the LBH model at a suitable energy scale,
as it will be shown in the next section.
\chapter{Renormalization Group Analysis}
\def\vplus{\gamma_+ e^{i\lambda\varphi}}
\def\vminus{\gamma_- e^{i\delta\eta-i\lambda\varphi}}
We want to consider here the renormalization group flow of the classical
Babelon-Bonora action
$$
S_{BB}=\intx\left[{{1\over 2}\kinx{\varphi}+\de_\mu\eta\de^\mu\xi
-2\left({e^{2\varphi}+e^{2\eta-2\varphi}}\right) }\right].
\eqn\bb
$$
At the quantum level one must implement wave and vertex function
renormalizations so that in \bb\ one must introduce different
bare coupling constants in front of the fields as well as in front of
the vertex interaction terms. As a consequence one ends with the form
\bblike. However, according to the general spirit of the
renormalization procedure we have also to consider Feigin--Fuchs terms
(the ones involving the $2D$-scalar curvature) since
all generally covariant dimension 2 counter terms are possible in \bblike.
This ansatz is in agreement with the perturbative theory as
one could show following Distler and Kawai.
In our context the Feigin--Fuchs terms come naturally out (see footnote
at page 6).
This leads us to consider the following
generalized form of the BB action in a curved space\rlap:\foot{We have
rotated $\varphi\to-i\varphi$, $\xi\to i\xi$ and $\eta\to-i\eta$ in order to
have the usual kinetic term. The model assumes in this way a form
analogous to the one of sine--Gordon (indeed it is also known as the
central sine--Gordon model)}
$$
\eqalign {
S &=\intx\sqrt{ g}\left[ {
g^{\mu\nu} \left({{1\over 2}\de_\mu\varphi\de_\nu\varphi +
\de_\mu\eta\de_\nu\xi}\right)
+\gamma_+e^{i\lambda\varphi}+\gamma_-e^{i\delta\eta-i\lambda\varphi}
}\right.\cr
&+\curv\varphi +\curv\eta +\curv\xi ] .\cr }
\eqn\gbb
$$
\par
We shall pursue here the renormalization procedure of
\gbb\ in a perturbative framework.
Notice that $\xi$ plays the role of an auxiliary field, a variation with
respect to which gives the on-shell equation of motion
$$
\nabla_\mu\nabla^\mu\eta=iq\xi R,
\eqn\consh
$$
which in our perturbative scheme must be linearized around the
flat space, giving:
$$
\de_\mu\de^\mu\eta=0.
\eqn\onsh
$$
This is the conservation law for the current $\de_\mu\eta$.
\REF\GLPZ{M.~T.~Grisaru, A.~Lerda, S.~Penati and D.~Zanon
\journal Nucl. Phys. &B342(90)564.}
Following Ref. \GLPZ, we define the renormalized quantities
at an arbitrary mass scale $\mu$ by:
$$
\matrix{
\eqalign{
\varphi &= Z^\unm_\varphi \varphi_R,\cr
\eta &= Z^\unm_\eta \eta_R,\cr
\xi &= Z^{-\unm}_\eta \xi_R,\cr}
&\eqalign{
\gamma_\pm &= \mu^2 Z_{\gamma_\pm} \gamma_{\pm R},\cr
\lambda^2 &= Z^{-1}_\varphi \lambda^2_R,\cr
\delta^2 &= Z^{-1}_\eta \delta^2_R,\cr}
&\eqalign{
 q^2_\varphi &= Z^{-1}_\varphi q^2_{\varphi R},\cr
 q^2_\eta &= Z^{-1}_\eta q^2_{\eta R},\cr
 q^2_\xi &= Z_\eta q^2_{\xi R}.\cr }
\cr}
\eqn\renquant
$$
The following quantities are conserved through renormalization:
$$
\matrix {
r={q_\varphi\over\lambda},
& k=q_\xi\delta,
& p=q_\xi q_\eta,\cr }
\eqn\constquant
$$
With our normalizations the regularized $\varphi\varphi$ or $\eta\xi$
propagator (the $\eta\eta$ and the $\xi\xi$ propagators are identically zero)
is:
$$
G(z-z') = -\unm\log\{ m_0^2 [(z-z')^2+\epsilon^2]\} ,
\eqn\renprop
$$
where $m_0$ and $\epsilon$ are an IR and an UV cutoff respectively.
Then we normal order the vertices, to eliminate tadpole
divergences, with the replacements:
$$
\eqalign {
e^{i\lambda\varphi} &\rightarrow (m_0^2\epsilon^2)^{\lambda^2\over 4}
:e^{i\lambda\varphi}:\cr
e^{-i\lambda\varphi+i\delta\eta} &\rightarrow
(m_0^2\epsilon^2)^{\lambda^2\over 4} :e^{-i\lambda\varphi+i\delta\eta}:\cr}
\eqn\vertrepl
$$
\par
Since $\lambda\varphi =\lambda_R\varphi_R$ and $\delta\eta =\delta_R\eta_R$,
the renormalization of the vertices is simply obtained by setting
$$
Z_{\gamma_\pm} = (\mu^2\epsilon^2)^{-\lambda^2/4},
\eqn\zetagammapm
$$
which gives the $\beta$ functions for the coupling constants
$\gamma$ in absence of curvature terms (\ie\ with the $q$'s set to zero):
$$
\beta_\pm := \dmu{\gamma_\pm} = 2\gamma_{\pm R}\left( {
{\lambda^2_R\over 4} -1}\right) ,
\eqn\betagpm
$$
so that in this case the ratio ${\gamma_+\over\gamma_-}$ is conserved
through the renormalization.
\par
As in Ref. \GLPZ\ we can calculate the field renormalizations considering
the average of the vertices $<V_+V_->$ (where $V_\pm$ are the exponential
interaction terms of \gbb) which, for $\lambda^2\sim 4$,
gives a contribution to the kinetic term of the form:
$$
-{1\over 4}\gamma_{+R}\gamma_{-R}\log (\mu^2\epsilon^2)\unm
(\lambda_R\de\varphi_R-\delta_R\de\eta_R)^2.
$$
Then the correct kinetic terms are obtained by putting:
$$
\eqalign {
Z_\varphi &= 1+{1\over 4}\gamma_{+R}\gamma_{-R}\lambda^2_R\log
\mu^2\epsilon^2,\cr
Z_\eta &= 1+{1\over 4}\gamma_{+R}\gamma_{-R}\delta^2_R\log
\mu^2\epsilon^2.\cr }
\eqn\zetape
$$
Notice that the renormalization has produced new terms proportional
to $\de_\mu\de^\mu\eta$, which however vanish if we consider the on-shell
quantum theory (\onsh) in flat space. The true on-shell theory should
rely on ~\consh, but at this perturbative order curvature terms
can be negelected (they are important, as we shall see, just in the
renormalization of $\gamma_\pm$).
Therefore we shall restrict ourself to
the on-shell renormalization scheme.
\par
We can now calculate all the remaining $\beta$ functions for $\lambda^2
\sim 4$:
$$
\eqalign {
\bettyr\lambda &= \boopr\lambda^4_R,\cr
\bettyr\delta &= \boopr\delta^4_R,\cr
\bettyr{{q_\varphi}} &= \boopr\lambda^2_R q^2_{\varphi R},\cr
\bettyr{{q_\eta}} &= \boopr\delta^2_R q^2_{\eta R},\cr
\bettyr{{q_\xi}} &= -\boopr\delta^2_R q^2_{\xi R}.\cr }
\eqn\betas
$$
\par
The task at this point is to
obtain the modifications to the $\beta$ functions due
to the curvature terms which are present in the quantum functional action.
For this purpose we first need the stress tensor of the theory:
$$
\eqalign {
T_{\mu\nu} &= 2\pi{2\over\sqrt { g}}
\left. {\delta S\over\delta g^{\mu\nu}} \right|_{ g=\eta}\cr
&= \de_\mu\varphi\de_\nu\varphi+2\de_\mu\eta\de_\nu\xi
-\eta_{\mu\nu}(\unm\de_\rho\varphi\de^\rho\varphi+\de_\rho\eta
\de^\rho\xi)+\cr
&-\eta_{\mu\nu}(\vplus+\vminus)+2i(\eta_{\mu\nu}\de_\rho\de^\rho-\de_\mu\de_\nu)
(q_\varphi\varphi+q_\eta\eta+q_\xi\xi),\cr }
\eqn\gbbst
$$
where $\eta_{\mu\nu}$ is the flat metric tensor.
Setting
$\gamma_\pm=0$ in \gbbst, we have the stress tensor of the kinetic part,
from which we can calculate the central charge of the free theory:
$$
c=3-24q_\varphi^2-48q_\xi q_\eta.
\eqn\pippo
$$
Notice that the total central charge, \ie\ the one involving also matter and
the ghosts contribution, is
$$
c_{tot}=c+N-26.
\eqn\ctot
$$
The trace of the stress tensor is easily calculated:
$$
T_\mu^{\ \mu} = -2[\vplus+\vminus+
2i\de_\mu\de^\mu (q_\varphi\varphi+q_\eta\eta+q_\xi\xi)],
\eqn\trgbbst
$$
and, using the classical equations of motion in flat space
$$
\eqalign {
\de_\mu\de^\mu\varphi &= i\lambda (\vplus -\vminus ),\cr
\de_\mu\de^\mu\eta &= 0,\cr
\de_\mu\de^\mu\xi &= i\delta\vminus ,\cr }
\eqn\cemf
$$
we find the classical expression:
$$
T_\mu^{\ \mu}=
-2[(1+q_\varphi\lambda)\vplus+(1-q_\varphi\lambda+q_\xi\delta)\vminus].
\eqn\ctrgbbst
$$
Following Zamolodchikov\rlap,\Ref\Zam{A.B.~Zamolodchikov
\journal Sov.~J.~Nucl. Phys. &46(87)1090.}
we obtain by ~\ctrgbbst\ the modified $\beta_\pm$ functions:
$$
\eqalign {
\beta_+ &= 2\gamma_+\left({{\lambda^2\over 4}-1-q_\varphi\lambda}\right),\cr
\beta_- &= 2\gamma_-\left({{\lambda^2\over 4}-1+q_\varphi\lambda-q_\xi\delta
}\right),\cr}
\eqn\mbetagpm
$$
while the others remain unchanged at this perturbative order.
{}From now on, for the sake of simplicity, we will omit the $R$ subscripts.
\par
Putting together the equations \betas, we find that also the following
quantity is conserved through renormalization\rlap:\foot{This result,
deriving from \betas, needs not be valid beyond this perturbative order,
while it is obviously true for the quantities in \constquant. Notice
moreover that putting $\lambda=\delta$ at an arbitrary scale sets $d=0$,
and since $d$ is a RG-invariant the condition $\lambda=\delta$ then holds
at any scale.}
$$
d={1\over\delta^2}-{1\over\lambda^2}.
\eqn\constd
$$
Using the non-perturbative RG-invariants in \constquant,
the RG equations for the $\gamma$'s can be simply rewritten as:
$$
\eqalign {
{d\over dt}\log(\gamma_+\gamma_-) &= \lambda^2-4-2k\cr
{d\over dt}\log{\gamma_-\over \gamma_+} &= 4r\lambda^2-2k,\cr}
\eqn\RGgamma
$$
where $t=\log\mu$.
\par
{}From the first of ~\RGgamma\ and the first of
{}~\betas\  in the approximate form $\beta_\lambda
\sim 8\gamma_+\gamma_-$, we easily obtain:
$$
{d\lambda^2\over dt}=\unm(\lambda^2-4-2k)^2-{b^2\over 2},
\eqn\RGlambda
$$
where $b$ is an integration constant supposed to be real in order to
have RG fixed points. These are located at:
$$
\lambda^2_\pm=4+2k\mp b,
\eqn\RGfix
$$
where, taking $b$ positive, $\lambda_+$ is the UV fixed point and $\lambda_-$
the IR fixed point.
{}~\RGlambda\ can now be solved in the form:
$$
t-t_0={1\over b}\log\left|{\lambda_-^2-\lambda^2\over\lambda^2-\lambda_+^2}
\right|,
\eqn\sollambda
$$
where $t_0$ is another integration constant.
Solving ~\sollambda\ with respect to $\lambda^2$ we
have:
$$
\lambda^2(t)={e^{b(t-t_0)}\lambda^2_+ +\lambda^2_-\over
e^{b(t-t_0)}+1}=
2k+4-b\tanh{b(t-t_0)\over 2}.
\eqn\sollambdain
$$
Now we can also easily integrate ~\RGgamma\ to get:
$$
\eqalign {
\gamma^2_- &=Ae^{(8rk+16r-2k)(t-t_0)}
\left[{\cosh{b(t-t_0)\over 2}}\right]^{-2-8r}\cr
\gamma^2_+ &= Be^{-(8rk+16r-2k)(t-t_0)}
\left[{\cosh{b(t-t_0)\over 2}}\right]^{-2+8r},\cr }
\eqn\solin
$$
where $A$ and $B$ are arbitrary constants. Notice that the first of
{}~\betas\ imposes
that $\gamma_+$ and $\gamma_-$ have opposite signs.
\par
So far we have described the most general situation in which all the
parameters are unconstrained. Indeed, following the reasoning of Sec.~2,
we have to request that at a certain scale $\hat t$, which shall be
proved to exist, both vertex operators have a conformal weight
$(1,1)$. This is achieved imposing the following
constraints\rlap:\foot{Notice that imposing these conditions is
equivalent to asking that, at the scale $\hat t$, both $\beta_\pm$ in
\betas\ vanish.}
$$
\eqalign{
\left({ {1\over 4}-r }\right)\hat\lambda^2&=1,\cr
\left({ {1\over 4}+r }\right)\hat\lambda^2&=k+1.\cr}
\eqn\constr
$$
Solving \constr\ we obtain:
$$
\eqalign{
\hat\lambda^2 &=2k+4={4\over1-4r},\cr
r &= {k\over4(2+k)}.\cr}
\eqn\BBsol
$$
This means by \sollambdain\ that the scale $\hat t$ is just the
renormalization scale $t_0$. At this scale we also require the vanishing
of the total central charge, ~\ctot, which gives:
$$
\eqalign{
c &= N-23-24r^2\lambda^2(t_0)-48p\cr
&= N-23 -24\left({
{4r^2\over1-4r}+2p}\right) =0.\cr}
\eqn\czero
$$
Notice moreover that we must have
$$
r<{1\over4},
\eqn\condreal
$$
if we want $\hat\lambda$ to be real.
It is also easily seen that the particular combination $8 rk+16
r-2k$ vanishes identically for any value of $k$, so that
$\gamma_\pm$ now read:
$$
\gamma_\pm(t)\propto\left[{
\cosh{b(t-t_0)\over2}}\right]^{-(1\mp4r)}.
\eqn\gammas
$$
This implies that $\gamma_\pm$ are even functions of
the scale $t-t_0$ and hence do not distinguish
between IR and UV scales. The requirement of having vertex operators
with the right conformal weight at the defining scale $t_0$ forces the theory
to be dual (under the exchange of the IR and UV scales).
The asymptotic form of $\gamma_-$ is
$$
\gamma_-\buildrel |t-t_0|\to\infty\over\sim
const.\ e^{-2s|t-t_0|},
\eqn\asyg
$$
where we have set
$$
s={1+4r\over4}|b|.
\eqn\defs
$$
\par
The next step in fixing the parameters is achieved considering what must
happen at the scale $t_{BC}$, where the theory becomes
the LBH model. This amounts to require that
the vertex operator $V_+$ must disappear at the scale
$t_{BC}$, \ie\ we must impose that
$$
\left|{{\gamma_-(t_{BC})\over\gamma_+(t_{BC})}}\right| >> 1.
\eqn\rap
$$
By ~\gammas, this requires that:
$$
r  < 0,
\eqn\cond
$$
and
$$
|b(t_{BC}-t_0)| >> 1.
\eqn\conduno
$$
{}From ~\conduno\ it follows that $\lambda(t_{BC})$ must be
equal to the asymptotic value $\lambda_\pm$ given by
eq. \RGfix.
\par
We first notice that as a consequence of the
rotation \rot, we have that
$$
\eqalign{
q_\varphi(t_{BC})&=G_{11}\sqrt{\kappa\over 2},\cr
q_\xi(t_{BC}) &=G_{12}\sqrt{\kappa\over 2},\cr
q_\eta(t_{BC}) &=G_{13}\sqrt{\kappa\over 2},\cr}
\eqn\qBC
$$
By \defld\ and \constquant\ we get:
$$
\eqalign{
r &={\kappa\over 4}{G_{11}\over G_{11}-{G_{21}+G_{31}\over\sqrt2}},\cr
k &=-2G_{12}\left({G_{13}-{G_{23}+G_{33}\over\sqrt2}}\right),\cr
p &={\kappa G_{12}G_{13}\over2}.\cr}
\eqn\rsol
$$
The condition $iii)$ of Sec. 2, together with ~\BBsol\ and \rsol\ imply that:
$$
G_{12}y^2-\kappa G_{11}G_{12}y-\kappa G_{11}=0,
\eqn\eqtre
$$
where we have put
$$
y=G_{11}-{G_{21}+G_{31}\over\sqrt2}.
\eqn\defx
$$
We have found a solution of \equno, \eqdue\ and \eqtre\ numerically.
We look for those solutions such that
\item{a)} $r$,
given by \rsol, is negative, in order to have the flow to the LBH model
and, at the same time, the consistency with ~\deuv;
\item {b)} $s$, given by ~\defs, is positive so that $\gamma_-$ is
bounded for large energy scale.
\par
A numerical solution satisfying the above conditions for $r$ and $s$
can be found for $-2<\kappa<0$ and $14<N<23$.
$\kappa$ is defined in ~\redoc. Its functional relation with the physical
parameter $N$ is defined implicitly in ~\czero and~\rsol.
To compute $s$ we also need the parameter
$b$, implicitly defined by ~\RGfix, which turns out to be:
$$
b=\lambda^2(t_{BC})-\lambda^2(t_0)=
{8\over\kappa}x^2-{4\over1-4r}.
\eqn\eqforb
$$
\chapter{Black Hole Thermodynamics}
In this section we want to discuss how our RG results affect the
black-hole thermodynamics.
Our strategy is to observe that at the scale $t_{BC}$ our theory reduces
to the LBH model which contains the simple black-hole solution of CGHS,
where the black-hole is formed by $N$ in-falling shock-waves. Now by replacing
the parameters of this solution (which we thus interpret as an "effective"
solution for our model with $u=v$) with those obtained with our RG analysis,
we can identify a temperature and the v.e.v. of the curvature.
The Hawking temperature
$T_H$ is proportional to $\sqrt{\gamma_-(t_{BC})}$:
$$
T_H(t)\propto\mu\sqrt{\gamma_-(t)},
\eqn\eqforTH
$$
which, by \asyg, goes asymptotically as
$$
T_H(t)\buildrel |t-t_0|\to\infty\over\sim
const.\ e^te^{-s|t-t_0|}.
\eqn\asy
$$
Using our solution  it is immediate to see
that there exists a dynamical regime for $23<N<30$
in which $T_H$ vanishes both in the UV and IR regions.
\par
The relation between the v.~e.~v. of the operator-valued
scalar curvature $\sqrt{-g} R$ and the other CATBH running coupling constant
$\lambda^{(-)}(t)$ in the
conformal gauge $\widehat{g_{\mu\nu}}=-\unm e^{2\lambda\rho}\eta_{\mu\nu}$
and in the tree approximation is:
$$
<\sqrt{-\hat g}\hat R> = 2\lambda^{(-)}(t)\de_\mu\de^\mu<\rho>.
\eqn\classeqforgR
$$
We may then state something about the end-point state of black hole
evaporation
if we use the CGHS-solution to describe the black
hole formation by $N$-shock waves $f_i$.
In our contest, the CGHS-ansatz for the classical solution $<\rho>\equiv
<\rho>_{\rm tree}$ looks:
$$
\eqalign{
e^{-2\lm(t)<\rho(x^+)>} &=
1-2\lm(t)<\rho(x^+)>+O(\lm^2)\cr
&=-\kappa a\theta(x^+-x_0^+)(x^+-x_0^+)-
e^{2t}\gamma_-(t)x^+x^-,\cr}
\eqn\sette
$$
where $x^\pm\equiv x^0\pm x^1$ and $a\equiv{\rm const}$.
Therefore, at $x^+=x^+_0$ we have by \classeqforgR, using the light-cone
coordinates $x^\pm$:
$$
\matrix{
<\sqrt{-\hat g(x^+)}\hat R(x^+)>=
\de_+\de_-[2\lm(t)<\rho(x^+)>]\sim e^{2t}\gamma_-(t),\hfill
&\hfill x^+\to x_0^+.\cr}
\eqn\otto
$$
Since here we have two scales $x^+_0$ and $\mu$, where $t\equiv\log (\mu)$,
it is reasonable to set (in $c=\hbar=1$) $x^+_0\equiv{1\over\mu}$, and
$\xop$ is
the natural scale which describes the classical
black hole formation.
Therefore:
$$
<[\radgR](e^{-t})>\propto e^{2t}\gamma_-(t).
\eqn\nove
$$
The Hawking temperature, \eqforTH, and the averaged density of the scalar
curvature, ~\nove, are asymptotically controlled by $\gamma_-(t)$.
\par
Since in the CGHS ansatz (Ref. \CGHS) the black hole mass $\mbh$ grows linearly
with $\xop$, \ie\ $\mbh\propto M^2_{\rm Pl}\xop$ where $M_{\rm Pl}$ is the
Planck mass, we get $\mbh\sim 1/\mu$. This relation may also be obtained
by a dimensional argument relying on Witten's relation\Ref\Wit{E. Witten
\journal Phys. Rev. &D44(91)314.} between the black hole mass and the
value $a$ of the dilaton field on the horizon (namely $\mbh\sim e^a$),
and, on the other hand, on the conformal properties of the
vertex $e^{-2\phi}$, which is recognized to be a primary field of
conformal (mass) dimension 2, so that, at a semiclassical level, we may
write $e^{-2a}\sim\mu^2$. Thus we get the previously stated relation
$\mbh\sim 1/\mu$.
\par
As a consequence of the above arguments, we get that ~\asy may be
rewritten as follows:
$$
T_H(\mbh)\buildrel\mbh\to0\over\sim T_0\left({\mbh\over m_0}\right)^{s-1},
\eqn\mtozero
$$
and
$$
T_H(\mbh)\buildrel\mbh\to\infty\over\sim
T_0'\left({m_0\over\mbh}\right)^{s+1},
\eqn\mtoinfty
$$
where $T_0,T_0'$ and $m_0$ are arbitrary constants.
The vanishing of the Hawking temperature for small and large black hole
masses occurs for $s>1$, which according to our numerical solution of
Sec. 3 requires $16<N<23$.
This is a consequence of the duality between the UV and IR limit of our
quantum theory.
In the following we
understand $N$ to be taken in the above range.
\par
The end point state of the black hole evaporation is characterized by the
limit $\mbh\to0$. But in this limit $T_H$ and by ~\nove\ also
$<\radgR>$ are vanishing. We understand this result
as a signal that at the end point the black hole \underbar{disappears
completely} from our $2D$-universe, leaving a zero temperature flat
remnant solution. This scenario has been suggested by
Hawking\refmark\Hawk and
't~Hooft\rlap,\Ref\tH{G. 't Hooft \journal Nucl. Phys. &B335(90)138.}
but with a basic difference: for Hawking ('t~Hooft) the final
state is a mixed (pure) state.
Of course at the level of the above RG analysis we
cannot say anything on the quantum black hole Hilbert space.
However, we have an explicit
quantum field model for answering, in principle, to the above question.
Our point of view is to see whether the $S$-matrix associated with
the ``quantum"
black hole states, which are in correspondence with the ``rotated"
Babelon-Bonora theory at the energy scale $t_{BC}$, \ie\ in terms
of the $u=v$ and
$\bar\chi$ fields, is unitary or not. Clearly a unitary $S$-matrix may be
in agreement only with 't~Hooft's scenario. In the following section,
we shall give some arguments which seem to support the \underbar{
non-unitarity picture}, and hence Hawking's point of view.
\chapter {Quantum Black Hole S-Matrix}
One starting point is the Babelon-Bonora version of our CATBH-model,
namely ~\bblike.
Using its Hopf
algebra structure, namely $\uq$, we shall get a quantum $S$-matrix and
then we shall pull it back to the physical black hole basis described by
the fields $u=v$ and $\chi$ of ~\redoc\ and \defex.
\par
The defining relations for the quantum Kac--Moody algebra $\uq$ are:
$$
\eqalign{
[H_i,H_j] &= 0,\cr
[H_i,E_j^\pm] &= \pm\al_i\cdot\al_j\ E_j^\pm,\cr
[E_i^+,E_j^-] &= \delta_{ij}{q^{H_i}-q^{-H_i}\over q-q^{-1}},\cr}
\eqn\suno
$$
where $i=0,1$; here $\al_0=-\al_1$ and $|\al_1|^2=2$. The center of $\uq$
is
$$
K = H_0 + H_1.
\eqn\sdue
$$
A new basis in $\uq$ is generated\foot{With CCR given by:
$$
\eqalign{
[H_i,Q_j] &= \al_i\cdot\al_j\ Q_j,\cr
[H_i,\Qb_j] &= -\al_i\cdot\al_j\ \Qb_j,\cr
Q_i\Qb_i-q^{-2}\Qb_iQ_i &={1-q^{2H_i}\over q^{-2}-1}.\cr}
$$} by $H_i,Q_i$ and $\Qb_i$:
$$
\eqalign{
Q_i &= E_i^+ q^{H_i\over 2},\cr
\Qb_i &= E_i^- q^{H_i\over 2}.\cr}
\eqn\stre
$$
\REF\RS{N.Y.
Reshetikhin and M.A. Semenon--Tian--Shansky\journal Lett. Math. Phys.
&19(90)133}
\REF\KT{B.M. Khoroshkin and V.N. Tolstoy, to apppear in {\sl Funkz.
Analyz. i ego pril.}}
The algebra $\uq$ is a qua\-si\-tri\-an\-gu\-lar Hopf
al\-ge\-bra\refmark\RS$^{^-}$\refmark\KT
with co\-mul\-ti\-pli\-ca\-tion
$$
\Delta:\uq\to\uq\otimes\uq
$$
defined by
$$
\eqalign{
\Delta(H_i) &= H_i\otimes1+1\otimes H_i,\cr
\Delta(Q_i) &= Q_i\otimes1+q^{H_i}\otimes Q_i,\cr
\Delta(\Qb_i) &= \Qb_i\otimes1+q^{H_i}\otimes \Qb_i.\cr}
\eqn\squattro
$$
The CAT model is associated to the $\uq$ Kac--Moody algebra by the
so-called homogeneous gradation:
$$
\matrix{
E_0^+ &= x^2\sigma_-,\hfill &\hfill E_0^- &=x^{-2}\sigma_+,\cr
E_1^+ &= \sigma_+,\hfill &\hfill E_1^- &= \sigma_-,\cr}
\eqn\scinque
$$
where $x$ is the ``spectral parameter" and the Pauli spin matrices
$\sigma_\pm$ are the usual step operators of $sl(2)$. Together with the
Pauli spin matrix $\sigma_3$, they form the so-called Chevalley basis for
$sl(2)$. The commutator in the loop algebra associated with our centered
$\widetilde{sl(2)}$ is defined as:
$$
\eqalign{
[A(x),B(x)] &= A(x)\cdot B(x)-B(x)\cdot A(x)+{1\over 2\pi i}\oint dx\ {\rm
tr} [\de_x A(x)\cdot B(x)]\ K,\cr
&= A(x)\cdot B(x)-B(x)\cdot A(x) + K(A,B).\cr}
\eqn\ssei
$$
The asymptotic soliton states are labelled by
$\ket{\tphi,\teta,\theta}$, where $\theta$ is the rapidity, $\tphi$
and $\teta$ are topological charges defined by:
$$
\eqalign{
x &=e^{\theta\left({{4\over\lambda^2}-1}\right)},\cr
\tphi &=\ {\lambda\over 2\pi}\int_{-\infty}^{+\infty}dx\ \de_x\varphi,\cr
\teta &=-{\lambda^3\over4\pi}\int_{-\infty}^{+\infty}dx\ \de_x\eta.\cr}
\eqn\ssette
$$
One has that
$$
\eqalign{
\tphi &= -H_0,\cr
\teta &= K,\cr}
\eqn\sotto
$$
and the relation with the $H_i$'s is the following
$$
\eqalign{
H_0 &= -\sigma_3+K,\cr
H_1 &= \sigma_3.\cr}
\eqn\snove
$$
The representation of $\uq$ on the space of one-soliton states can be
shown to be:
$$
\matrix{
Q_+ &= cQ_1 &= c\sigma_+q^{\sigma_3\over2},\hfill
&\hfill Q_- &= cQ_0 &= cx^2\sigma_-q^{-\sigma_3+K\over2},\hfill\cr
\Qb_- &= c\Qb_1 &= c\sigma_-q^{\sigma_3\over2},\hfill
&\hfill \Qb_+ &= c\Qb_0 &= cx^{-2}\sigma_+q^{-\sigma_3+K\over2},\hfill\cr}
\eqn\sdieci
$$
where $c$ is a constant depending linearly on $\gamma$
and the deformation parameter $q$ is given by:
$$
q=e^{4\pi i\over\lambda^2}.
\eqn\dodici
$$
The two-soliton to two-soliton $S$-matrix $\stilde$ is an operator from
$V_1\otimes V_2$ to $V_2\otimes V_1$, $V_i$ are the vector spaces spanned
by $\ket{\tphi,\teta,\theta}$. The $S$-matrix must
commute\Ref\BL{D.~Bernard and A. LeClair\journal Comm. Math. Phys.
&142(91)99} with the
action of $\uq$, since it is the symmetry group of the theory:
$$
[\stilde,\Delta(H_i)]=[\stilde,\Delta(Q_\pm)]=
[\stilde,\Delta(\Qb_\pm)]=0.
\eqn\stredici
$$
The representation of eq.~\stredici\ on $V_1\otimes V_2$ is explicitly
given by:
$$
[\stilde,\sigma_3\otimes1+1\otimes\sigma_3]=0,
\eqn\squattordicia
$$
$$
\eqalign{
&\stilde\cdot(\sigma_+q^{\sigma_3\over2}\otimes1+q^{\sigma_3}\otimes\sigma_+
q^{\sigma_3\over2})+K(\stilde,\Delta(Q_+)) =\cr
&(\sigma_+q^{\sigma_3\over2}\otimes1+q^{\sigma_3}\otimes\sigma_+
q^{\sigma_3\over2})\cdot\stilde,\cr
&\stilde\cdot(x_1^2\sigma_-q^{-\sigma_3+K_1\over2}\otimes1+q^{-\sigma_3+K_1}
\otimes x_2^2\sigma_-q^{-\sigma_3+K_2\over2}) +K(\stilde,\Delta(Q_-))=\cr
&(x_2^2\sigma_-q^{-\sigma_3+K_2\over2}\otimes1+q^{-\sigma_3+K_2}
\otimes x_1^2\sigma_-q^{-\sigma_3+K_1\over2})\cdot\stilde,\cr}
\eqn\squattordicib
$$
$$
\eqalign{
&\stilde\cdot(x_1^{-2}\sigma_+q^{-\sigma_3+K_1\over2}\otimes1
+q^{-\sigma_3+K_1}\otimes x_2^{-2}\sigma_+q^{-\sigma_3+K_2\over2})
+K(\stilde,\Delta(\Qb_+))=\cr
&(x_2^{-2}\sigma_+q^{-\sigma_3+K_2\over2}\otimes1
+q^{-\sigma_3+K_2}\otimes x_1^{-2}\sigma_+q^{-\sigma_3+K_1\over2})
\cdot\stilde,\cr
&\stilde\cdot(\sigma_-q^{\sigma_3\over2}\otimes1+q^{\sigma_3}\otimes\sigma_-
q^{\sigma_3\over2})+K(\stilde,\Delta(\Qb_-))=\cr
&(\sigma_-q^{\sigma_3\over2}\otimes1+q^{\sigma_3}\otimes\sigma_-
q^{\sigma_3\over2})\cdot\stilde.\cr}
\eqn\squattordicic
$$
A solution $\stilde=\stilde(x_1/x_2,K_1,K_2,q)$ of eq.~\stredici\ is of the
following form\rlap:\Ref\Jim{M. Jimbo \journal Comm. Math. Phys.
& 102(86)537.}
$$
\stilde=f(x_1/x_2,q,K_1,K_2)R(x_1/x_2,q,K_1,K_2),
\eqn\squindici
$$
where $R$ is the universal quantum $R$-matrix associated to $\uq$, whose
existence has been proved in Ref.~\RS. One can easily show that $R$
satisfies the quantum Yang--Baxter equation, which is required for the
factorization of the multisoliton $S$-matrix:
$$
R_{12}(x)R_{13}(xy)R_{23}(y)=R_{23}(y)R_{13}(xy)R_{12}(x).
\eqn\ssedici
$$
\REF\Ros{M. Rosso \journal Comm. Math. Phys. &124(89)307.}
\REF\KR{A.N. Kirillov and N. Reshetikhin, {\it q-Weyl Group and
Multiplicative Formula for Universal R-Matrices}, preprint HUTMP
90/B261.}
An explicit form of the universal $S$-matrix for $\uq$ is given in Ref.
\Ros\ and \KR. In the ``bootstrap" approach
the overall factor $f$ can be found by imposing crossing and
unitarity conditions. Of course, this is consistent if the CAT-model
belongs to the class of $2D$ relativistic quantum field theories studied
by Zamolodchikov and Zamolodchikov\rlap.\Ref\ZZ{A.B. Zamolodchikov and
A.B. Zamolodchikov \journal Ann. Phys. & 120(79)253.} However in our
context it is not necessary to answer to this question and to find an
explicit form for $f$, since the relevant two-body scattering matrix is
the one, denoted by $\sbh$,
acting on the black hole basis $\ket{\chi,u=v}$. We shall see below
that, even if a unitary $S$-matrix in the CAT basis could be found,
$\sbh$ is not.
Formally $\sbh$ is defined by
$$
\sbh=U^{\dag} \stilde U,
\eqn\sdiciassette
$$
where $U=PG$, and $G$ is the operator associated to the rotation
\rot\ and $P$ is the projection operator from $\ket{\chi,u,v}$ onto
the black hole basis $\ket{\chi,u=v}$. It is evident that $\sbh$ is no
more an automorphism of the Hilbert space $\cal H$
spanned by the CAT-basis and
hence, for a well-known theorem\rlap,
\Ref\barry{J.Weidmann,{\it Linear Operators in Hilbert Space},
Springer-Verlag (New York,1980).}
is not an unitary operator on $\cal H$.
The conclusion that we draw for the black hole evaporation scenario is
that the final state is approached incoherently, even if full quantum
gravity effects are taken into account, supporting Hawking's point of
view.
\ack
L. Bonora and A. Ashtekar are thanked for some illuminating discussions.
\endpage\refout
\bye